\documentclass[12pt]{article}
\usepackage{amsmath}

\usepackage{accents,amssymb,amsthm,color,enumerate,fancyhdr,geometry,hyperref}
\allowdisplaybreaks[4]
\usepackage[all,pdf]{xy}
\usepackage{pifont}
\usepackage[perpage,symbol*]{footmisc}
\usepackage{cases}
\usepackage{graphicx}
\usepackage{subcaption}
\usepackage{appendix}
\usepackage{cite}
\usepackage{bbm}
\usepackage{bm}
\usepackage{lineno} 
\usepackage{changepage} 

\usepackage[T1]{fontenc}
\usepackage{authblk}
\usepackage{natbib}
\setcitestyle{authoryear,round}

\usepackage{amsfonts}
\usepackage{float}
\usepackage{abstract}

\usepackage{microtype}
\usepackage{booktabs} % for professional tables
\usepackage{multirow}
\usepackage{makecell}
\usepackage[figuresright]{rotating}
\usepackage{array}
\usepackage[export]{adjustbox}
\usepackage{anyfontsize}
\usepackage{enumitem}
\usepackage{algorithm}
\usepackage{caption}
\usepackage{longtable}

\usepackage{algpseudocode}

\usepackage{hyperref}
\hypersetup{hypertex = true,
colorlinks = true,
linkcolor = blue,
anchorcolor = blue,
citecolor = blue}

\usepackage{setspace,fullpage}

\title{\textbf{A Multilayer Probit Network Model for Community Detection with Dependent Edges and Layers}}

\author[1]{Dapeng Shi}
\author[2]{Haoran Zhang}
\author[3]{Tiandong Wang}
\author[1]{Junhui Wang}

\affil[1]{Department of Statistics,
 The Chinese University of Hong Kong}
\affil[2]{Department of Statistics and Data Science, Southern University of Science and Technology}
\affil[3]{Shanghai Center for Mathematical Sciences, Fudan University}
 
\date{}

\newcommand{\boA}{\boldsymbol{A}}

\newcommand{\bomu}{\boldsymbol{\mu}}
\newcommand{\boSigma}{\boldsymbol{\Sigma}}
\newcommand{\boTheta}{\boldsymbol{\Theta}}

\newcommand{\boe}{\boldsymbol{e}}
\newcommand{\bou}{\boldsymbol{u}}

\newcommand{\boZ}{\boldsymbol{Z}}

\newcommand{\boOmega}{\boldsymbol{\Omega} }
\newcommand{\boomega}{\boldsymbol{\omega} }

\newcommand{\bbP}{\mathbb{P}}
\newcommand{\bbE}{\mathbb{E}}

\newcommand{\argmin}{\operatorname*{argmin}} % define the argmin operator
\newcommand{\argmax}{\operatorname*{argmax}} % define the argmin operator

\newtheorem{assumption}{Assumption}
\newtheorem{lemma}{Lemma}
\newtheorem{theorem}{Theorem}
\newtheorem{corollary}{Corollary}
\newtheorem{proposition}{Proposition}

\newcommand\keywords[1]{\textbf{Keywords}: #1}

\begin{document}

%\pagewiselinenumbers	

\maketitle

\onehalfspacing
\begin{abstract}	
Community detection in multilayer networks, which aims to identify groups of nodes exhibiting similar connectivity patterns across multiple network layers, has attracted considerable attention in recent years. Most existing methods are based on the assumption that different layers are either independent or follow specific dependence structures,
and edges within the same layer are independent. In this article, we propose a novel method for community detection in multilayer networks that accounts for a broad range of inter-layer and intra-layer dependence structures. The proposed method integrates the multilayer stochastic block model for community detection with a multivariate probit model to capture the structures of inter-layer dependence, which also allows intra-layer dependence.
To facilitate parameter estimation, we develop a constrained pairwise likelihood method coupled with an efficient alternating updating algorithm. The asymptotic properties of the proposed method are also established, with a focus on examining the influence of inter-layer and intra-layer dependences on the accuracy of both parameter estimation and community detection. The theoretical results are supported by extensive numerical experiments on both simulated networks and a real-world multilayer trade network.
\end{abstract}

\doublespacing
\section{Introduction}

Multilayer networks are a powerful representation of relational data, where nodes represent entities and edges across different layers capture multiple relationships between those entities. These networks have been increasingly encountered in various real applications, such as biological networks \citep{lei2020consistent, lei2023bias}, trade networks \citep{jing2021community, jiang2023autoregressive}, and social networks \citep{han2015consistent}. 

In multilayer networks, the detection of homogeneous community structure in different layers has been extensively studied \citep{lei2020consistent,jing2021community}. Most existing community detection methods for multilayer networks are adaptations of those designed for single-layer networks, such as matrix decomposition-based or spectral-based methods \citep{paul2020spectral,tang2009clustering,kumar2011co}, likelihood-based methods \citep{han2015consistent,paul2016consistent,Yuan2021CD}, and least squares-based methods \citep{lei2020consistent,lei2023bias}.
These methods are mostly based on the assumption that the network layers are mutually independent.

Nevertheless, multilayer networks often exhibit various dependence structures between and within network layers. For instance, in time-varying networks, autoregressive correlations are commonly found between adjacent networks \citep{jiang2023autoregressive, Zhang01102020,Athreya03042025,chatterjee2022}; in trade networks, goods with similar attributes tend to have strong correlations due to shared market dynamics and trade partners, while goods with different attributes show weaker or even ignorable correlations \citep{jing2021community,Lyu02102023}; in spatial networks, geographically closer entities typically exhibit stronger correlations, driven by shared local markets or regional policies \citep{Liuglobal}; in brain networks, the same subject is often measured multiple times, leading to correlations among repeated brain connectivity networks \citep{Konstantinos}. These various types of dependence structures highlight the importance of incorporating such dependencies for more accurate and efficient network analysis. 

Only very recently were inter-layer and intra-layer dependence incorporated into multilayer network modeling. For example, \citet{jiang2023autoregressive} introduced the autoregressive stochastic block model (SBM) to capture inter-layer dependence with a time series structure; \citet{zhang2024} proposed the multilayer Ising model to capture the full inter-layer dependence; \citet{Yuan2021CD} proposed the multilayer SBM that incorporates the dependence within the community to account for the intra-layer dependence. However, it remains unclear how to extend \citet{jiang2023autoregressive} to accommodate more general dependence structures, while the method in \citet{zhang2024} faces challenges in estimating connection probabilities due to the intractable computational cost of the partition function \citep{ravikumar2010high}. Additionally, \citet{Yuan2021CD} requires the assumption of layer independence conditional on additional covariates to consistently estimate the correlations between edges. Moreover, very little has been done in the literature to theoretically investigate the impact of dependence structures on the accuracy of community detection.

In this article, we introduce a multilayer probit network model that integrates the classical multilayer SBM \citep{lei2020consistent,holland1983stochastic} with the multivariate probit model \citep{heagerty1998composite}, allowing both inter-layer and intra-layer dependences in network modeling. Originally developed for spatially correlated responses, the multivariate probit model is well suited to capture various dependence structures. The proposed multilayer probit network model also relates to the latent Gaussian copula model for binary data \citep{fan2017high}, by introducing Gaussian latent factors to model discrete data. To facilitate parameter estimation and community detection, we propose to utilize a constrained pairwise likelihood 
%as a surrogate objective 
and develop an efficient alternative updating scheme to tackle the resulting optimization task. 

Theoretically, we establish the asymptotic consistency of the proposed method for both parameter estimation and community detection. More importantly, we demonstrate how the inter-layer and intra-layer dependences affect the accuracy of community detection. Through extensive simulations and a real-world multilayer trade network, we demonstrate the superior numerical performance of the proposed method compared to several popular competitors.

The remainder of the article is organized as follows. Section \ref{Sec:model} introduces the proposed model. In Section \ref{Sec:Pairlike}, we propose a constrained pairwise likelihood function along with an efficient alternating update algorithm to address the resulting optimization problem. Section \ref{Sec:asymptoics} establishes the asymptotic consistency of the proposed method for both parameter estimation and community detection. In Section \ref{Sec:experiments}, we evaluate the performance of our approach through simulated and real-world examples, comparing it against several popular competing methods in literature. Finally, Section \ref{Sec:discuss} offers a brief summary. All technical proofs are provided in the appendix and the supplementary materials.

\paragraph{Notations.} We summarize the notation used throughout this paper. For an integer $K$, let $S_K$ denote the set of all permutations of $[K]$.
Given two sequences of real numbers $\left\{a_N\right\}$ and $\left\{b_N\right\}, a_N=O\left(b_N\right)$,  $a_N \lesssim b_N$ and $b_N\gtrsim a_N$ represent that $\left|a_N\right| \leq c_1\left|b_N\right|$ for a constant $c_1>0$, $a_N=o\left(b_N\right)$  and $ b_N\gg a_N $ mean $\lim _{N \rightarrow \infty} a_N / b_N=0$, and $a_n \asymp b_N$ means $c_2\left|b_N\right|\leq\left|a_N\right| \leq c_3\left|b_N\right|$ for constants $c_2,c_3>0$. Given a sequence of random variables $\left\{X_N\right\}$, we write $X_N=O_p\left(a_N\right)$ if for any $\epsilon>0$, there exists finite $C>0$ such that $\sup _N \bbP\left(\left|X_N / a_N\right|>C\right)<\epsilon$. For a vector $v=(v^{(1)},\cdots, v^{(M)})\in\mathbb{R}^M$, let $
 \|v\|=\sqrt{\sum_{b=1}^M(v^{(b)})^2}$ be the Euclidean norm.  For a matrix $W=\left(w^{(bd)}\right)_{N\times M} \in \mathbb{R}^{M\times M}$, let $\|W\|=\sqrt{\sum_{b=1}^N\sum_{d=1}^M (w^{(bd)})^2}$, $\|W\|_{\max} = \max_{1\leq b\leq N, 1\leq d\leq M} |w^{(bd)}|$, $\operatorname{Supp}(W) = \{ (b,d): w^{(bd)}\not= 0 \}$, $|W| = |\operatorname{Supp}(W)|$, $W_i$ be the $i$-th row of $W$ and $W_{-i}$ be the submatrix of $W$ excluding its $i$-th row. In addition, for a matrix $Y\in\mathbb{R}^{N\times N}$, the operator $\operatorname{diag}(Y) $ refers to the diagonal matrix formed by the diagonal elements of $Y$,  while $\operatorname{ndiag}(Y) = Y - \operatorname{diag}(Y)$. Moreover, $R \succ 0$ indicates that $R$ is positive definite in $\mathbb{R}^{M\times M}$,  meaning that for any non-zero vector 
$v\in \mathbb{R}^M$, $v^\top Rv > 0$.  Finally, denote  $\mathbf{1}_M$ as the vector in $\mathbb{R}^M$ with all elements being 1.

%{[define notations here]} $\asymp$, $\gtrsim,\lesssim$,$\gg$ , $\|\cdot\|_{\max}, \|\cdot\|$, $\operatorname{diag}(X)$, $\operatorname{ndiag}(X)$,$X_{N} = O(Y_N)$, $X_{N} = o(Y_N)$, $X_{N} = O_P(Y_N)$, ,$\mathbf{1}_N $, let $Z_{-i}$ as the submatrix of $Z$ excluding its $i$-th row,

\section{Multilayer probit network}\label{Sec:model}

Let $\mathcal{G}$ denote a multilayer network comprising $M$ network layers on $N$ common nodes, where each network layer can be represented via its adjacency matrix $\boA^{(b)} = (A^{(b)}_{ij})_{N \times N} \in \{0, 1\}^{N \times N}$ for $b\in[M]$. Here, $A^{(b)}_{ij} = A^{(b)}_{ji} = 1$ if an edge exists between nodes $i$ and $j$ in the $b$-th layer, and $A^{(b)}_{ij} = A^{(b)}_{ji} = 0$ otherwise. Let $\boldsymbol{\varepsilon}_{ij} =  \big(\varepsilon_{ij}^{(1)}, \cdots, \varepsilon_{ij}^{(M)} \big)^\top \in\mathbb{R}^{M}$, and  we consider a multilayer probit network model, 
\begin{align*}
& A^{(b)}_{ij} = \mathbb{I} \big \{ \mu^{(b)}_{e_ie_j} +  \varepsilon_{ij}^{(b)} > 0 \big \}, \ \mbox{for any } b \in [M], \\
& \boldsymbol{\varepsilon}_{ij}\sim N \big (0, \boSigma_{e_ie_j} \big ), \ \mbox{for any } i,j \in [N], 
\end{align*}
where $\mathbb{I}(\cdot)$ is the indicator function, $e_i \in [K]$ denotes the homogeneous community membership of node $i$ across $M$ layers, $\bomu^{(b)} = (\mu_{kl}^{(b)})_{k,l=1}^K \in \mathbb{R}^{K \times K}$ denotes the mean matrix for each network layer $b$, and $\boSigma_{kl} = (\Sigma_{kl}^{(bd)})_{b,d\in[M]}$ denotes the positive definite covariance matrix for each $k,l\in[K]$ to accommodate dependence among layers. We allow the inter-layer dependence structures to vary across communities. Further assume that $$
\operatorname{corr}\big(\boldsymbol{\varepsilon}_{i_1j_1}, \boldsymbol{\varepsilon}_{i_2j_2}\big) = \operatorname{diag}(r^{(1)}_{i_1j_1, i_2j_2}, \cdots, r^{(M)}_{i_1j_1, i_2j_2}),
$$ 
with $r^{(b)}_{i_1j_1, i_2j_2} \in [-1,1]$ accommodating dependence among edges within the same layer. Note that the marginal distribution of $\boA^{(b)}$ follows an SBM model with the mean matrix $ \big(\Phi(\mu^{(b)}_{kl})\big)_{k,l} $, where $\Phi(\cdot)$ is the cumulative distribution function of the standard normal distribution. Also, the correlations between each pair of edges may differ, which generalizes \citet{Yuan2021CD}, who assumes independence between edges across different communities. Similar edge dependence modeling has also been considered in the beta model for single layer network \citep{yan2018probitnetworkmodelarbitrary}.

It is evident that both layer dependence and edge dependence are accommodated in the proposed multilayer probit network model. Define the edge index set
$D_{kl}=\{(i,j):\,e_i=k,\ e_j=l,\ i<j\}$ for $k,l\in [K]$,  and  $n_{kl}=|D_{kl}|$ such that $\sum_{k, l\in[K]} n_{kl} = N(N-1)/2$.
Stack edge-wise within each type
$
\widetilde{\boldsymbol{\varepsilon}}_{kl}
=\big(\boldsymbol{\varepsilon}_{ij}^\top\big)_{(i,j)\in D_{kl}}^\top
\in\mathbb{R}^{n_{kl} M},
$
and then stack all types $k,l\in[K]$  to form
$
\boldsymbol{\varepsilon}
=\big(\widetilde{\boldsymbol{\varepsilon}}_{11}^\top,\cdots, \widetilde{\boldsymbol{\varepsilon}}_{KK}^\top\big)^\top
\in\mathbb{R}^{MN(N-1)/2}.
$ 
Define a block-diagonal matrix 
$
\boldsymbol{S}
=\mathrm{diag}\big \{\boldsymbol{I}_{n_{11}}\otimes\boldsymbol{\Sigma}_{11}^{1/2},
\cdots,
\boldsymbol{I}_{n_{KK}}\otimes\boldsymbol{\Sigma}_{KK}^{1/2}\big \} \in\mathbb{R}^{(MN(N-1)/2)\times(MN(N-1)/2)},
$
and the edge-level correlation matrix $\boldsymbol{P}\in\mathbb{R}^{MN(N-1)/2\times MN(N-1)/2}$,
partitioned by types as
$\boldsymbol{P}=\big[\boldsymbol{P}_{k_1l_1,k_2l_2}\big]_{k_1,k_2, l_1,l_2\in [K]}$ with $\boldsymbol{P}_{k_1l_1,k_2l_2}\in\mathbb{R}^{Mn_{k_1l_1}\times Mn_{k_2l_2}}$ being the correlation matrix between edges in $D_{k_1l_1}$ and $D_{k_2l_2}$ across all layers. In particular, $\boldsymbol{P}_{k_1l_1,k_2l_2} = [ \boldsymbol{P}_{k_1l_1,k_2l_2, i_1j_1, i_2j_2}]_{\substack{(i_1,j_1)\in D_{k_1l_1}\\ (i_2,j_2)\in D_{k_2l_2}}} $ with $ \boldsymbol{P}_{k_1l_1,k_2l_2, i_1j_1, i_2j_2}= \operatorname{diag}(r^{(1)}_{i_1j_1,i_2 j_2},\cdots,r^{(M)}_{i_1j_1,i_2 j_2})\in\mathbb{R}^{M\times M} $. Then the error distribution can be rewritten as $\boldsymbol{\varepsilon}\sim N \big(\mathbf{0},\boldsymbol{S}\boldsymbol{P}\boldsymbol{S}\big)$. Note that when $r_{i_1j_1,i_2j_2}^{(b)} = r_{i_1j_1,i_2j_2}$ for $b\in[M]$, which means the correlation between edges $(i_1,j_1)$ and $(i_2,j_2)$ remains the same across all layers,
the error distribution can be simplified as $\boldsymbol{\varepsilon}\sim N \big(\mathbf{0},\boldsymbol{S}(\boldsymbol{R}\otimes \boldsymbol{I}_M)\boldsymbol{S}\big)$ with $\boldsymbol{R} = [r_{i_1j_1,i_2j_2} ]_{i_1<j_1,i_2<j_2\in [N]} \in \mathbb{R}^{N(N-1)/2\times N(N-1)/2}$.

Note that the dependence structure between any pair of network layers $ \boA^{(b)} $ and $ \boA^{(d)} $ can be effectively captured by $ \boSigma_{kl} $, owing to the properties of the multivariate probit model. Specifically, when $ e_i = k, e_j = l $ and $ b \neq d $, $ A_{ij}^{(b)} $ and $ A_{ij}^{(d)} $ are independent if $ \Sigma_{kl}^{(bd)} = 0 $. In addition, if $ \Sigma_{kl}^{(bd)} > 0 $, then $ A_{ij}^{(b)} $ and $ A_{ij}^{(d)} $ are positively correlated; conversely, if $ \Sigma_{kl}^{(bd)} < 0 $, they are negatively correlated. For model identifiability, we further assume that all diagonal elements of $\boSigma_{kl}$ are equal to 1, while the off-diagonal elements can be flexibly modeled to accommodate various dependence structures among network layers. 
In fact, by varying the off-diagonal elements of $\boSigma_{kl}$, the multilayer probit network model includes many standard network models as special cases when there exists no edge dependence within the same network layer. For instance, it degenerates to the multilayer SBM \citep{lei2020consistent} when all $\boSigma_{kl}$'s are identity matrices; it also degenerates to the autoregressive SBM \citep{jiang2023autoregressive} when only the diagonals and first off-diagonals in $(\boSigma_{kl}^{-1})$'s are non-zero; it becomes equivalent to the multilayer Ising model \citep{zhang2024} when all elements in $\boSigma_{kl}$'s are non-zero. It is clear that the proposed multilayer probit model allows various inter-layer dependence structures among the network layers by formulating $\boSigma_{kl}$'s properly. 

Furthermore, the proposed model only involves $O(K^2M + \sum_{k,l}|\boSigma_{kl}|)$ parameters to capture the connectivity probabilities and inter-layer dependence structures, which is substantially fewer than those required by the multivariate Bernoulli distribution \citep{chandna2022edgecoherencemultiplexnetworks,song2023independencetestinginhomogeneousrandom}. In particular, when $\boSigma_{kl}$ is sparse with $|\boSigma_{kl}| \lesssim M$, the inter-layer dependence structures do not increase the model complexity even when $M$ is large. In addition, the proposed model contains $O(N^4M)$ parameters, $r_{i_1j_1,i_2j_2}^{(b)}$ for $i_1,i_2,j_1,j_2\in[N]$ and $b\in[M]$, to capture the edge dependence within the same network layer, most of which may be zero and we will quantify the effect of the edge dependence sparsity to the estimation accuracy in Section~\ref{Sec:asymptoics}.

Let $\bomu = (\mu_{kl}^{(b)})_{k,l\in[K],b\in[M]}$
and $\boSigma = (\Sigma_{kl}^{(bd)})_{k,l\in[K],b,d\in[M]}$.
Denote $\boTheta = (\bomu,\boSigma)$ with $\boTheta_{kl}^{(bd)} = (\mu_{kl}^{(b)},\mu_{kl}^{(d)},\Sigma_{kl}^{(bd)})$, and $\boZ = (Z_{ik})_{i \in [N]; k \in [K]}$ as the homogeneous community membership matrix, where $Z_{ik} = 1$ if $e_i = k$, and $Z_{ik} = 0$ otherwise. The full likelihood of the multilayer network can be written as
\begin{equation}\label{eq:full_lik}
    \mathcal{L}_{\operatorname{full}}(\boTheta, \{r_{i_1j_1,i_2j_2}^{(b)}\}, \boZ) := 
    \frac{1}{(2\pi)^{MN(N-1) / 4} \, |\boldsymbol{S} \boldsymbol{P} \boldsymbol{S}|^{1/2}} \int_{\bou \in \boldsymbol{U}} \exp\left( -\frac{1}{2} \boldsymbol{u}^\top \left( \boldsymbol{S} \boldsymbol{P} \boldsymbol{S} \right)^{-1} \boldsymbol{u} \right)d\bou,
\end{equation}
where $\bou = \left(\bou_{11}, \cdots, \bou_{KK}\right)$, $\boldsymbol{U} = \boldsymbol{U}_{11}\times\cdots\times \boldsymbol{U}_{KK} $, $\boldsymbol{U}_{kl} = \times_{ \{(i,j):Z_{ik} = 1, Z_{jl} = 1\}} U_{ij}, U_{ij} = U_{ij,kl}^{(1)} \times \cdots \times U_{ij,kl}^{(M)} $, with $U_{ij,kl}^{(b)} = \big[\mu_{kl}^{(b)}, \infty\big)$ if $A_{ij}^{(b)} = 0$, and $U_{ij,kl}^{(b)} = \big(-\infty, \mu_{kl}^{(b)}\big]$ otherwise. The multi-dimensional integrals in \eqref{eq:full_lik} can be computationally inefficient, which would require substantial efforts to facilitate parameter estimation.

%Moreover, the number of edge dependence parameters $\{r_{i_1j_1,i_2j_2}^{(b)}\}_{b\in[M],i_1,i_2,j_1,j_2\in [N]}$ is $O(N^4 M)$, which exceeds the total sample size of the multilayer network data, making it impossible to estimate these parameters accurately.

\section{Pairwise likelihood estimation}\label{Sec:Pairlike}

Instead of the full likelihood, we consider a pairwise likelihood function as an alternative, which ignores the edge dependence parameters $\{r_{i_1j_1,i_2j_2}^{(b)}\}$ to facilitate computation \citep{varin2011overview}. Specifically, we replace \eqref{eq:full_lik} with
\[
\prod_{k,l\in[K]}\prod_{1\leq i<j\leq N}\prod_{1 \leq b < d \leq M} \bbP\big( A_{ij}^{(b)}, A_{ij}^{(d)}; \boTheta^{(bd)}_{kl} \big),
\] 
where 
\[
\begin{aligned}
\bbP\big( A_{ij}^{(b)}, A_{ij}^{(d)}; \boTheta^{(bd)}_{kl} \big) &= \alpha_1\left( \boTheta^{(bd)}_{kl} \right)^{A_{ij}^{(b)} A_{ij}^{(d)}} \times \alpha_2\left( \boTheta^{(bd)}_{kl} \right)^{A_{ij}^{(b)} (1 - A_{ij}^{(d)})} \\
&\quad \times \alpha_3\left( \boTheta^{(bd)}_{kl} \right)^{(1 - A_{ij}^{(b)}) A_{ij}^{(d)}} \times \alpha_4\left( \boTheta^{(bd)}_{kl} \right)^{(1 - A_{ij}^{(b)}) (1 - A_{ij}^{(d)})}.
\end{aligned}
\]
The terms $\alpha_1, \alpha_2, \alpha_3$ and $\alpha_4$ are defined as
\begin{equation}\label{eq:prob}
\begin{aligned}
\alpha_1\big( \boTheta^{(bd)}_{kl} \big) &= \bbP\big( A_{ij}^{(b)} = 1, A_{ij}^{(d)} = 1 ; \boTheta^{(bd)}_{kl} \big) = \Phi_2\big(\mu_{kl}^{(b)},\mu_{kl}^{(d)}, \Sigma^{(bd)}_{kl} \big), \\
\alpha_2\big( \boTheta^{(bd)}_{kl} \big) &= \bbP\big( A_{ij}^{(b)} = 1, A_{ij}^{(d)} = 0 ; \boTheta^{(bd)}_{kl} \big) = \Phi\big(\mu_{kl}^{(b)}\big) -\Phi_2\big(\mu_{kl}^{(b)},\mu_{kl}^{(d)}, \Sigma^{(bd)}_{kl} \big), \\
\alpha_3\big( \boTheta^{(bd)}_{kl} \big) &= \bbP\big( A_{ij}^{(b)} = 0, A_{ij}^{(d)} = 1 ; \boTheta^{(bd)}_{kl} \big) = \Phi\big(\mu_{kl}^{(d)}\big) - \Phi_2\big(\mu_{kl}^{(b)},\mu_{kl}^{(d)}, \Sigma^{(bd)}_{kl} \big), \\
\alpha_4\big( \boTheta^{(bd)}_{kl} \big) &= \bbP\big( A_{ij}^{(b)} = 0, A_{ij}^{(d)} = 0 ; \boTheta^{(bd)}_{kl} \big) = 1 -\Phi\big(\mu_{kl}^{(b)}\big)-\Phi\big(\mu_{kl}^{(b)}\big)+\Phi_2\big(\mu_{kl}^{(b)},\mu_{kl}^{(d)}, \Sigma^{(bd)}_{kl} \big),
\end{aligned}
\end{equation}
where $\Phi(\cdot)$ is the cumulative distribution function of $N(0,1)$, and $\Phi_2(\cdot, \cdot, \sigma)$ is the cumulative distribution function of 
$N_2 \big (( \begin{smallmatrix} 0 \\ 0 \end{smallmatrix} ), \big( \begin{smallmatrix} 1 & \sigma \\ \sigma & 1 \end{smallmatrix} \big) \big )$. The pairwise log-likelihood then becomes
\begin{align*}
    \mathcal{L} (\boTheta, \boZ) =& \sum_{k,l\in[K]}\sum_{i<j}\sum_{b<d} Z_{ik}Z_{jl} \Big \{A_{ij}^{(b)}A_{ij}^{(d)}\log \alpha_1 ( \boTheta^{(bd)}_{kl} ) +  A_{ij}^{(b)}(1-A_{ij}^{(d)})\log\alpha_2 ( \boTheta^{(bd)}_{kl}) \\
    & +(1 - A_{ij}^{(b)})A_{ij}^{(d)}\log \alpha_3\left( \boTheta^{(bd)}_{kl}\right) +(1-A_{ij}^{(b)})(1 - A_{ij}^{(d)})\log \alpha_4 ( \boTheta^{(bd)}_{kl}) \Big \} \\
    =&: \sum_{k,l}\mathcal{L}_{kl} (\boTheta, \boZ ).
\end{align*}

Denote $\mathcal{S}_{kl} $ as the pre-specified, shape-constrained set for $\boSigma_{kl}$. Specifically, we focus on two scenarios, the sparse covariance matrix scenario with $$
\mathcal{S}_{kl} = \left\{ \boldsymbol{X} \in \mathbb{R}^{M \times M} \mid \boldsymbol{X} \succ 0, \operatorname{diag}(\boldsymbol{X}) = \mathbf{1}_M, \operatorname{Supp}(\boldsymbol{X}) = T_{kl} \right\},
$$ and the sparse precision matrix scenario with
$$
\mathcal{S}_{kl} = \left\{ \boldsymbol{X} \in \mathbb{R}^{M \times M} \mid \boldsymbol{X} \succ 0, \operatorname{diag}(\boldsymbol{X}) = \mathbf{1}_M, \operatorname{Supp}(\boldsymbol{X}^{-1}) = T_{kl} \right\}.
$$ In both cases, $ T_{kl} \subseteq [M] \times [M] $ represents the set of positions, known a priori, with $ |T_{kl}| = s_{kl}^*+M $. Two examples for each scenario are the multilayer Ising model \citep{zhang2024} and the autoregressive SBM \citep{jiang2023autoregressive}. Define the parameter space as
\begin{align*}
\boOmega = \bigg\{ \boomega = (\bomu, \boSigma,  \boZ) \mid \boZ\in\{0,1\}^{N\times K},~ \boZ\mathbf{1}_K = \mathbf{1}_N,~ 
c_l\rho_{N,M} \leq \Phi(\mu_{kl}^{(b)}) \leq c_u \rho_{N,M}, \\
 \boSigma_{kl} \in \mathcal{S}_{kl},~\text{and}~ \sup_{k,l} \| \operatorname{ndiag}(\boSigma_{kl}) \|_{\max} \leq D_{N,M} \bigg\},
\end{align*}
where $c_l < 1 < c_u$ are two constants and $\rho_{N,M}$ controls the network sparsity level which shrinks to zero no faster than $\frac{\log(NM)}{NM}$. %We require all network layers share similar network sparsity.
Moreover, we impose the condition $D_{N,M} < 1$ to avoid singularities when deriving the pairwise likelihood terms in \eqref{eq:prob}. 
Note that the magnitudes of $s_{kl}^*$ and $D_{N,M}$ specify the inter-layer dependence structures and the strength of dependence across different layers, respectively.  Denote the true values of parameters as $\boomega^* = (\boTheta^*, \boZ^*) = (\bomu^*, \boSigma^*, \boZ^*)$ and  $\boldsymbol{P}^*$,  and assume $\boomega^*\in \boOmega$.  Lemma \ref{Lemma.Fisherconsis} shows that the pairwise likelihood function in \eqref{eq:Likelihood} is Fisher consistent in $\boOmega$.

\begin{lemma}\label{Lemma.Fisherconsis}
Let $\boe (\boomega^*, \boomega ) = \frac{1}{N^2M^2} \sum_{k,l} \bbE \big ( \mathcal{L}_{kl} (\boTheta^*, \boZ^* ) - \mathcal{L}_{kl}\left(\boTheta, \boZ \right)\big )$, then it holds true that $\boe(\boomega^*, \boomega) \geq 0$ for any $\boomega \in \boOmega$.
\end{lemma}

Lemma \ref{Lemma.Fisherconsis} shows that $\boomega^*$ is a maximizer of $\bbE(\mathcal{L}(\boTheta, \boZ))$, and thus justifies the use of the pairwise likelihood function in estimating $\boomega^*$.
Therefore, we estimate $(\boTheta^*,\boZ^*)$ via the constrained maximum pairwise log-likelihood estimate,
\begin{align}\label{eq:Likelihood}
  (\widehat{\boTheta}, \widehat{\boZ}) = \argmax_{(\boTheta,\boZ)\in\boOmega}\mathcal{L}\left(\boTheta, \boZ\right).  
\end{align}
We adopt a similar alternative updating algorithm as in \citet{Yuan2021CD} and \citet{zhang2024}. Given $\boTheta^{(t)}$ and $\boldsymbol{Z}^{(t)}$ at the $t$-th step, we first update $\boTheta = (\bomu, \boSigma)$ using projected gradient ascent. Specifically, for each $1 \leq k \leq l \leq K$, define $\bomu_{kl} = (\mu_{kl}^{(b)})\in\mathbb{R}^M$ and $$
\begin{aligned}
    (\bomu_{kl})^{(t, s+1)} &= \mathcal{P}_{\mathcal{U}_{kl}} \Big ( (\bomu_{kl})^{(t, s)} + \beta^{(t, s)} \frac{\partial \mathcal{L}_{kl} (\boTheta^{(t, s)}, \mathbf{Z}^{(t)})}{\partial \bomu_{kl}} \Big ), \\
    (\boSigma_{kl})^{(t, s+1)} &= \mathcal{P}_{\mathcal{S}_{kl}} \Big ( (\boSigma_{kl})^{(t, s)} + \beta^{(t,s)} \frac{\partial \mathcal{L}_{kl}(\boTheta^{(t, s)}, \mathbf{Z}^{(t)})}{\partial \boSigma_{kl}} \Big ),
\end{aligned}
$$
where $t, s \geq 0$, $\beta^{(t, s)} > 0$ denotes the step size, which can be adaptively selected using the line search method,  and $\boTheta^{(t,0)}$ is simply initialized as $\boTheta^{(t)}$.
In addition, $\mathcal{P}_{\mathcal{U}_{kl}}, \mathcal{P}_{\mathcal{S}_{kl}}(\cdot)$ represent the projection operators onto the spaces $\mathcal{U}_{kl} := \{ \boldsymbol{u} = (u^{(1)}, \dots, u^{(M)}) \in \mathbb{R}^{M} \mid c_l \rho_{N,M} \leq \Phi(u^{(b)}) \leq c_u \rho_{N,M} \}$ and $\mathcal{S}_{kl}$ under the Euclidean norm, respectively.  The details about each term in $\frac{\mathcal{L}_{kl}(\boTheta^{(t, s)}, \mathbf{Z}^{(t)})}{\partial \bomu_{kl}}$ and $\frac{\partial \mathcal{L}_{kl}(\boTheta^{(t, s)}, \mathbf{Z}^{(t)})}{\partial \boSigma_{kl}}$ are deferred to the supplementary materials. For each $1 \leq k \leq l \leq K$, we update $((\bomu_{kl})^{(t, s)}, (\boSigma_{kl})^{(t, s)})$ until the relative error of $\mathcal{L}_{kl}(\boTheta^{(t, s)}, \boZ^{(t)})$, defined as
$
\frac{\left| \mathcal{L}_{kl}\left(\boTheta^{(t, s+1)}, \boZ^{(t)}\right) - \mathcal{L}_{kl}\left(\boTheta^{(t, s)}, \boZ^{(t)}\right) \right|}{\left|\mathcal{L}_{kl}\left(\boTheta^{(t, s)}, \boZ^{(t)}\right)\right|}
$
is smaller than some predefined threshold $\delta_1$, or the number of updating steps exceeds $T_{1,\max}$. We further define the final update $\boTheta^{(t,s+1)}$ as $\boTheta^{(t+1,0)}$.

Optimizing $\mathcal{L}(\boTheta, \boZ)$ over $\boZ$ is computationally expensive due to its discrete nature. Following the idea of \citet{Yuan2021CD} and \citet{zhang2024}, we assign node $i$ to the $k$-th community if
\[
\mathcal{L}\left(\boTheta, \boZ_{-i}, Z_{ik}=1\right) \geq \mathcal{L}\left(\boTheta, \boZ_{-i}, Z_{il}=1\right), \ \text{for all } l \neq k.
\]
Thus, for each $i \in [N]$ and $k \in [K]$, we update $Z_{ik}$ as follows:
\[
Z_{ik}^{(t+1)} = \left\{
\begin{array}{l}
1, \ \ \text{when } k = \argmax_{k \in [K]}\mathcal{L}\left(\boTheta^{(t+1)}, \boZ
_{-i}^{(t)}, Z_{ik}=1\right), \\
0, \ \ \text{otherwise}.
\end{array}
\right.
\]
Note that $\boZ^{(t)}$ can be updated efficiently since $\mathcal{L}\left(\boTheta^{(t+1)}, \boZ
_{-i}^{(t)}, Z_{ik}=1\right)$ is derived from $\mathcal{L}\left(\boTheta^{(t+1)}, \boZ^{(t)}\right)$ by considering only the terms related to $\boldsymbol{Z}_i^{(t)}$, and thus the update can be performed in parallel for all $\boZ_i$.

The whole updating procedure terminates when
$
\frac{\left|\mathcal{L}\left(\boTheta^{(t+1)}, \boZ^{(t+1)}\right) - \mathcal{L}\left(\boTheta^{(t)}, \boZ^{(t)}\right)\right|}{\left|\mathcal{L}\left(\boTheta^{(t)}, \boZ^{(t)}\right)\right|} < \delta_2
$
or when the number of updating steps exceeds $T_{2,\max}$. The final estimates are denoted as $(\widehat{\boTheta}, \widehat{\boldsymbol{Z}})$. Note that the developed algorithm only converges to some local maximum \citep{Yuan2021CD}, thus it shall be beneficial to consider multiple initializations, despite the increased computational cost.  In the implementation of the proposed algorithm, we initialize $\boZ$ using the estimated community memberships from \citet{lei2020consistent} or \citet{paul2020spectral}, and $\boTheta$ with multiple random warm-up runs, which leads to satisfactory numerical performance. The complete algorithm is summarized in Algorithm \ref{Algo}.

\begin{algorithm}
\caption{Iterative optimization for updating \(\boTheta\) and \(\boZ\)}
\begin{algorithmic}[1]
\State \textbf{Input:} Stopping error: $\delta_1,\delta_2$. Maximum iteration steps: $T_{1,\max}, T_{2,\max}$. 
\State \textbf{Initialization:} Let $\boTheta^{(0)}$ be the identity matrix and \(\boZ^{(0)}\) be the estimated community memberships from \citet{lei2020consistent}.
\While{the relative error $\frac{\left| \mathcal{L}(\boTheta^{(t+1)}, \boZ^{(t+1)}) - \mathcal{L}(\boTheta^{(t)}, \boZ^{(t)}) \right|}{|\mathcal{L}(\boTheta^{(t)}, \boZ^{(t)})|} > \delta_2$ or $t\leq T_{2,\max}$ }  
\While{the relative error $\frac{\left| \mathcal{L}_{kl}\left(\boTheta^{(t, s+1)}, \boZ^{(t)}\right) - \mathcal{L}_{kl}\left(\boTheta^{(t, s)}, \boZ^{(t)}\right) \right|}{\left|\mathcal{L}_{kl}\left(\boTheta^{(t, s)}, \boZ^{(t)}\right)\right|} >\delta_1  $ or $s\leq T_{1,\max}$ }
    \For{each \(k,l \in [K]\)} 
       \State Update \(\bomu_{kl}\) and \(\boSigma_{kl}\) using projected gradient ascent:
            \[
            (\bomu_{kl})^{(t,s+1)} = \mathcal{P}_{\mathcal{U}_{kl}} \left( (\bomu_{kl})^{(t,s)} + \beta^{(t,s)} \frac{\partial \mathcal{L}_{kl}}{\partial \bomu_{kl}} \right)
            \]
            \[
            (\boSigma_{kl})^{(t,s+1)} = \mathcal{P}_{\mathcal{S}_{kl}} \left( (\boSigma_{kl})^{(t,s)} + \beta^{(t,s)} \frac{\partial \mathcal{L}_{kl}}{\partial \boSigma_{kl}} \right)
            \]
    \EndFor
   \EndWhile
   \State Set \(\boTheta^{(t+1)} = \boTheta^{(t+1,0)} = \boTheta^{(t,s+1)}\)
    \State Update \(\boZ^{(t+1)}\):
    \[
    Z_{ik}^{(t+1)} = \left\{
    \begin{array}{ll}
    1, & \text{if } k = \arg\max \mathcal{L}\left(\boTheta^{(t+1)}, \boZ_{-i}^{(t)}, Z_{ik}=1\right) \\
    0, & \text{otherwise}
    \end{array}
    \right.
    \]
    \EndWhile
\State \textbf{Output:} The last iteration values $(\boTheta^{\operatorname{final}},\boZ^{\operatorname{final}})$.
\end{algorithmic}
\label{Algo}
\end{algorithm}

As computational remarks, the proposed method requires knowledge of $\operatorname{Supp}(\boSigma_{kl}^*)$ or $\operatorname{Supp}(\boSigma_{kl}^{*-1})$. When $\operatorname{Supp}(\boSigma_{kl}^*)$ is unknown, we adopt the following four-stage procedure. 
First, for each layer pair $b<d$, we solve the optimization problem in \eqref{eq:Likelihood} based on the two-layer network $(\boA^{(b)}, \boA^{(d)})$ and obtain the initial estimates $\{\Sigma_{kl}^{(0),(bd)}\}_{k,l\in[K]}$ and $\widehat{\boZ}^{(bd)}$. 
Let $\widehat{\mathcal{C}}_k^{(bd)} = \{i : \widehat{Z}_{ik}^{(bd)} = 1\}$ be the $k$-th estimated community estimated from networks $(\boA^{(b)},\boA^{(d)})$. 
%and $(\pi_1^{(bd)},\cdots,\pi_K^{(bd)})$ be the permutation of $[K]$. 
Then, for each $b<d$, we solve the following optimization problem to align the community memberships obtained from different layer pairs to the first two layers: 
$$
\begin{aligned}
(\pi_1^{(1),(bd)},\cdots,\pi_K^{(1),(bd)}) = \argmin_{(\pi_1,\cdots,\pi_K)\in S_K} \frac{1}{N}\sum_{k=1}^K\big|\widehat{\mathcal{C}}^{(12)}_k\setminus\widehat{\mathcal{C}}_{\pi_k}^{(bd)}\big|.
\end{aligned}
$$
Next, we apply the element-wise hard-thresholding operator of \citet{Rothman01032009} to construct an initial estimate of $\operatorname{Supp}(\boSigma_{kl}^*)$ as $\big\{(b,d)\in[M]^2:|\Sigma^{(0),(bd)}_{\pi_k^{(1),(bd)}\pi_l^{(1),(bd)}}| > \lambda_{N,M}\big\}$ for some pre-speficed thresholding $\lambda_{N,M}$.
Finally, this support estimate is incorporated into $\mathcal{S}_{kl}$, and we resolve the optimization problem in \eqref{eq:Likelihood} to obtain the final estimators $(\widetilde{\boTheta}, \widetilde{\boZ})$. The complete algorithm and the theoretical guarantees for support recovery are deferred to Appendix B. Moreover, since the exact sparsity level $\rho_{N,M}$ of the networks is unknown in practice, we approximate it by the empirical quantity $
\widehat{\rho}_{N,M} = \frac{2}{N(N-1)M} \sum_{b \in [M]} \sum_{i<j \in [N]} A_{ij}^{(b)}.
$

%define $\boSigma_{kl}^{(1)} = (\Sigma^{(0),(bd)}_{\pi_k^{(1),(bd)}\pi_l^{(1),(bd)}})_{b,d\in [M]} $.  $$
%\Sigma^{(2),(bd)}_{kl} = \Sigma^{(1),(bd)}_{kl} \,\mathbb{I}\bigl\{ \lvert \Sigma^{(1),(bd)}_{kl} \rvert > \lambda_{N,M} \bigr\},
%$$ to construct an initial estimate of $\operatorname{Supp}(\boSigma_{kl}^*)$. 

%When $\boSigma_{kl}^{*-1}$ is unknown, we introduce an additional optimization step as $(\boSigma_{kl}^{-1})^{(0)} = \argmin_{\boldsymbol{X}\succ 0} \|\boldsymbol{X}\boSigma_{kl}^{(0)} - I\|_F$, and then apply the element-wise hard thresholding procedure to $(\boSigma_{kl}^{-1})^{(0)}$ to obtain an initial estimate of $\operatorname{Supp}(\boSigma_{kl}^{*-1})$.

\section{Asymptotics}\label{Sec:asymptoics}

We now investigate the asymptotic consistency of the proposed method, which is referred to as MLP for simplicity, in terms of both parameter estimation and community detection.

\subsection{Consistency in parameter estimation}
To quantify the effect of the edge dependence structure within each layer on the accuracy of parameter estimation, we need to introduce two additional quantities. 
Let $\Gamma = \{(i,j)\in[N]^2: i<j\}$ denote the edge set, and we say some subset $\mathcal{A}\subseteq \Gamma$ is independent if the corresponding variables $\{\boA_{ij}\}_{(i,j)\in\mathcal{A}}$
are independent. Define $\chi_{N}$ as the smallest number of disjoint independent subsets needed to cover the set $\Gamma$.
%size of the smallest proper cover of $\Gamma$, which is the smallest number of disjoint independent subsets needed to cover the set $\Gamma$ completely. The formal definition is deferred to Appendix D.
Then we can partition $\Gamma$ into a union of disjoint subsets $J_1, \dots, J_{\chi_{N}}$, where $\{\boA_{ij}\}_{(i,j) \in J_l}$ are mutually independent within each subset $J_l$. Further, define the quantity 
$$
\kappa_N = \max_{l \in [\chi_{N}]} \left( \frac{|J_l|}{\min_j |\{i : (i,j) \in J_l\}|} + \frac{|J_l|}{\min_i |\{j : (i,j) \in J_l\}|} \right), 
$$
which will appear in the upper bound of the convergence rate. Note that when $J_l$ is a triangle, meaning that $J_l = \{(i,j)\in[N]^2:~ i<j,~a_1\leq i\leq a_2,~b_1\leq j\leq b_2 \}$ for some $a_1<a_2$ and $b_1<b_2$, then $\frac{|J_l|}{\min_j |\{i : (i,j) \in J_l\}|}$ and $\frac{|J_l|}{\min_i |\{j : (i,j) \in J_l\}|}$ equal $b_2-b_1$ and $a_2-a_1$, respectively.
We illustrate the notions of $\chi_{N}$ and $\kappa_N$ with following three examples. 
\begin{enumerate}
    \item If all $\boA_{ij}$ are completely dependent, then $\chi_{N} = N^2$ and $\kappa_N \asymp 1$.
    
    \item When each edge is only correlated with a bounded number of edges, meaning that there exists a constant $c$ such that for any edge $\boA_{ij}$, $\left|\{(u,v): \operatorname{corr}(\boA_{ij}, \boA_{uv}) \neq 0\}\right| \leq c$, then $\chi_{N} \asymp 1$ and $\kappa_N \asymp N$. 

    \item For networks with hub structures \citep{Yuan2021CD}, where two edges are dependent if and only if they share a common node, we have $\chi_{N} \asymp N$ and $\kappa_N \asymp N$.
\end{enumerate}
We further provide an upper bound for $\chi_{N}$ for the general dependence structure between edges. Define the edgewise %second-order pairwise 
correlation density as
\begin{align*}
\tau_{N,M} 
= \max_{i,j}\frac{\Big|\{  (u,v): \text{there exists} \ b\in[M] \ \text{such that} \ r_{ij,uv}^{(b)}\not= 0     \}\Big|}{N(N-1)/2 - 1},
\end{align*}
where $N(N-1)/2 - 1$ is the number of edges. It follows from \citet{janson2004large} that $\chi_{N} \lesssim N^2 \tau_{N,M}$. 
%and thus we can replace $\chi_{N}$ by $N^2 \tau_{N,M}$ in Proposition \ref{Pro.errorlike} to simplify the result.

%The second example considers a scenario where all $\boA_{ij}$ are completely dependent leading to $\chi_{N} = N^2$ and $\kappa_N \asymp 1$. The third example focuses on networks with hub structures \citep{Yuan2021CD}, where two edges are correlated if they share a common node, and are independent otherwise. In this case, it can be verified that $\chi_{N} \asymp N$ and $\kappa_N \asymp N$.

The following Proposition~\ref{Pro.errorlike} quantifies the convergence rate of the estimation error in terms of the measure $e(\boomega, \boomega^*)$, which is the basis for deriving the convergence rate in terms of more common measure such as Hellinger distance.

%Assumption \ref{Assum.Sigma} holds, 
\begin{proposition}\label{Pro.errorlike}
Suppose that $\frac{1}{1-D_{N,M}} = o\big(\Phi^{-1}(\rho_{N,M})\big)$. Then there exist a sequence $\{\varepsilon_{N,M}\}$ and some constant $C$ such that
\begin{equation}
\label{eq:epsilon}
1 \geq \varepsilon_{N,M}^2 \geq  C   \frac{\chi_{N}\Big(\kappa_N K  + K^2M+\sum_{k,l}s^*_{kl}\Big)B^2_{N,M}}{N^2M(1-D_{N,M})^2  }\log\frac{KN M}{\varepsilon_{N,M}(1-D_{N,M}) } ,
\end{equation}
with $B_{N,M} := (\Phi^{-1}(\rho_{N,M}))^2$, when $N$ and $M$ are sufficiently large. Further, it holds true that
\[
\begin{aligned}
&\bbP\bigg(\sup _{\{\boomega \in \boOmega: \boe(\boomega^*, \boomega) \geq \varepsilon_{N,M}^2\}} \big (\mathcal{L}(\boomega) - \mathcal{L} (\boomega^* ) \big ) \geq 0 \bigg) 
\leq  2\exp\Big (- \frac{C_1B_{N,M}}{1-D_{N,M}}\log \Big ( \frac{KNM}{1-D_{N,M}}\Big ) \Big ),
\end{aligned}
\]
for some positive constant $C_1$.
\end{proposition}

Proposition \ref{Pro.errorlike} ensures that any $\boomega$ satisfying $\mathcal{L}(\boomega) \ge \mathcal{L} (\boomega^*)$ lies in the neighborhood of $\boomega^*$ with high probability. 
When each edge is only correlated with a bounded number of edges, then $\varepsilon_{N,M}^2$ in \eqref{eq:epsilon} is governed by the total number of observations $N^2M$, the total number of parameters to be estimated $ NK + K^2M + \sum_{k,l}s_{kl}^*$, and the strength of dependence $D_{N,M}$. 
Particularly, when $K \lesssim \log(NM)$,  $M \lesssim \sqrt{N}$, and $1 - D_{N,M}\gtrsim \frac{1}{\sqrt{\log(NM)}} $, Proposition \ref{Pro.errorlike} implies that $\varepsilon_{N,M}^2 \gtrsim \frac{1}{NM}$  up to some logarithmic terms, even when layers are mutually dependent such that $\sum_{k,l}s_{kl}^* \asymp M^2$. This demonstrates that the estimated parameters converge to the true ones at a fast rate, regardless of the dependence structure. In addition, for the homogeneous SBM \citep{Yuan2021CD}, where $\mu_{kl}^{(b)} = \mu_{kl}$ for $b\in [M]$, and $r_{i_1j_1, i_2j_2}^{(b)} = 0$ if either $e_{i_1}\neq e_{i_2}$ or $e_{j_1}\neq e_{j_2}$, then $\chi_{N} \asymp N^2$ and $\kappa_N \asymp K$. If further $K \lesssim \log(NM)$ and $\sum_{k,l} s_{kl}^* \lesssim M$, Proposition \ref{Pro.errorlike} implies that $\varepsilon_{N,M}^2 \gtrsim \frac{1}{M}$, up to logarithmic terms, which aligns with the rate of the independent likelihood method proposed by \citet{Yuan2021CD}.

Denote the pairwise likelihood estimate as $\widehat{\boomega}$, and its estimation error is measured by the averaged Hellinger distance,
\[
h^2(\widehat{\boomega}, \boomega^*) = \frac{1}{N^2M^2}\sum_{i,j} \sum_{b<d} h^2(\widehat{\boTheta}_{\widehat{e}_i\widehat{e}_j}^{(bd)}, \boTheta^{*, (bd)}_{e^*_i e^*_j}),
\]
where $h(\widehat{\boTheta}_{\widehat{e}_i\widehat{e}_j}^{(bd)}, \boTheta^{*, (bd)}_{e^*_i e^*_j})$ is the discrete Hellinger distance between $\alpha_c\big( \widehat{\boTheta}_{\widehat{e}_i\widehat{e}_j}^{(bd)} \big)$ and $\alpha_c\big( \boTheta^{*, (bd)}_{e^*_i e^*_j}\big)$ for $c = 1, \dots, 4$ as
\[
h^2(\widehat{\boTheta}_{\widehat{e}_i\widehat{e}_j}^{(bd)}, \boTheta^{*, (bd)}_{e^*_i e^*_j}) = \sum_{c=1}^4 \Big (\sqrt{\alpha_c\big( \widehat{\boTheta}_{\widehat{e}_i\widehat{e}_j}^{(bd)}\big)} - \sqrt{\alpha_c\big( \boTheta^{*,(bd)}_{ e^*_i e^*_j}\big)} \Big )^2.
\] Proposition \ref{Pro.errorlike} provides a crucial intermediate result for analyzing the convergence rate of $\widehat{\boTheta}$, by establishing a connection between $\boe(\boomega^*, \boomega)$ and $h(\widehat{\boomega},\boomega^*)$. The following Theorem~\ref{Thm.pararate} further gives the convergence rate of $\widehat\boomega$ to $\boomega^*$ in terms of the averaged Hellinger distance.

\begin{theorem}\label{Thm.pararate}
Suppose all assumptions in Proposition \ref{Pro.errorlike} hold. Then, 
%there exists a local maximizer of $\mathcal{L}(\boTheta, \boldsymbol{Z})$, denoted as $\widehat{\boomega}$, such that
\[
\bbP( h(\widehat{\boomega}, \boomega^*) \geq \varepsilon_{N,M} ) \leq 2\exp\Big (- \frac{C_1B_{N,M}}{1-D_{N,M}}\log \Big ( \frac{KNM}{1-D_{N,M}}\Big ) \Big ),
\]
implying that $h(\widehat{\boomega}, \boomega^*) = O_p(\varepsilon_{N,M})$. Moreover,
\[
    \frac{1}{MN^2} \sum_{i,j}\sum_{b} \big ( \widehat{\mu}_{\widehat{e}_i\widehat{e}_j}^{(b)} - \mu_{e_i^*e_j^*}^{*,(b)} \big )^2 +
\frac{(1-D_{N,M})B_{N,M}}{M^2N^2} \sum_{i,j}\sum_{b<d} \big (\widehat{\Sigma}^{(bd)}_{\widehat{e}_i\widehat{e}_j} - \Sigma^{*,(bd)}_{e_i^*e_j^*} \big )^2 = O_p\Big(\frac{\varepsilon_{N,M}^2}{B_{N,M}\rho_{N,M}}\Big),
\]
where $\widehat{e}_i$ denotes the estimated community membership of node $i$ such that $Z_{i\widehat{e}_i} = 1$. % for any $i \in [N]$.
\end{theorem}

Theorem \ref{Thm.pararate} establishes that $\widehat\boomega$ converge to $\boomega^*$ at a fast rate, which depends on the network size $N$, the total number of layers $M$, the inter-layer dependence structures $s_{kl}^*$, the egde dependence structures $\chi_{N}$ and $\kappa_N$, the strength of inter-layer  dependence $D_{N,M}$, and the network sparsity factor $\rho_{N,M}$. Particularly, the consistency result in Theorem \ref{Thm.pararate} holds true as long as $\rho_{N,M} \gg \varepsilon_{N,M}^2$. When each edge is only correlated with a bounded number of edges, for $K \lesssim \log(NM)$,  $M \lesssim \sqrt{N}$ and any dependence structure $s_{kl}^*$, then $\rho_{N,M}$ can be of order $\frac{1}{NM}$ up to some logarithmic terms, which matches the best sparsity condition in \citet{paul2020spectral} for independent multilayer networks. For networks with hub structure such that two edges are dependent if and only if they share a common node, when $K \lesssim \log(NM)$, $\sum_{k,l} s_{kl}^* \lesssim M$, the required sparsity condition is $\rho_{N,M} \gg \tfrac{1}{M} + \frac{1}{N}$, up to logarithmic factors. In this context, consistency is influenced by the increase in the number of either the nodes or the network layers.

\subsection{Consistency in community detection}

Let $\mathcal{C}_k^* = \{i : Z_{ik}^* = 1\}$ be the set of nodes belonging to the $k$-th true community, and $\widehat{\mathcal{C}}_k = \{i : \widehat{Z}_{ik} = 1\}$ be the $k$-th estimated community. Then the accuracy of community detection is evaluated as
\begin{equation}\label{eq:distZ}
  \operatorname{dist}(\widehat{\boldsymbol{Z}},\boldsymbol{Z}^*) = \min_{(\pi_1, \cdots, \pi_K) \in S_K} \frac{1}{N} \sum_{k \in [K]} \big|\mathcal{C}_k^* \setminus \widehat{\mathcal{C}}_{\pi_k}\big|, 
\end{equation}
where recall that $|\cdot|$ represents the cardinality of a set, $\mathcal{C}_k^* \setminus \widehat{\mathcal{C}}_{\pi_k}$ contains nodes in $\mathcal{C}_k^*$ but not in $\widehat{\mathcal{C}}_{\pi_k}$. Note that this definition also accounts for the fact that community labels are invariant to a permutation. The following two assumptions are required for the consistency of community detection.

\begin{assumption}\label{Assum.Comgap}
There exist a constant $C_2>0$ and a positive sequence $\{\eta_{N,M}\}$ such that
\[
\min_{b, k_1 \neq k_2} \frac{1}{\sqrt{K}}\big \|\bomu_{k_1}^{*,(b)} - \bomu_{k_2}^{*,(b)} \big\| \geq C_2\eta_{N,M},
\] 
when $N$ and $M$ are sufficiently large, where $\bomu_{k}^{*,(b)} = (\mu^{(b)}_{k1}, \cdots, \mu^{(b)}_{kK})$ for each $k \in [K]$.
\end{assumption}

\begin{assumption}\label{Assum.Combalance}
There exists a constant $C_3>0$ such that $\min_{k \in [K]} |\mathcal{C}_k^*| \geq C_3 \frac{N}{K}$ when $N$ is sufficiently large.
\end{assumption}

Assumption \ref{Assum.Comgap} is an identifiability condition, which requires that the true communities remain sufficiently separated. This condition is critical for ensuring the feasibility of community detection \citep{ZhenHyper,zhang2024}. The degree of separation is quantified by the positive sequence $\eta_{N,M}$, which may decrease to 0 at a slow rate. Assumption \ref{Assum.Combalance} guarantees that the communities are well-defined and do not degenerate too fast asymptotically. This is a mild condition that is typically satisfied when vertex community memberships are drawn from a multinomial distribution. Similar assumptions have been proposed in \citet{ke2019community} for community detection in undirected hypergraphs, whereas \citet{chien2019minimax} adopts a stricter condition, assuming that community sizes are equal.

\begin{theorem}\label{Thm.Comconsist}
Suppose all assumptions in Theorem \ref{Thm.pararate} and Assumptions \ref{Assum.Comgap} - \ref{Assum.Combalance} hold. Then, it holds true that
\[
\operatorname{dist}(\widehat{\boZ},\boZ^*) = O_p\Big (\frac{K  \varepsilon_{N,M}^2}{\eta_{N,M}^{2}B_{N,M}\rho_{N,M}}\Big ),
\]
provided that $\frac{K  \varepsilon_{N,M}^2}{\eta_{N,M}^{2}B_{N,M}\rho_{N,M}} =o(1)$, where $\varepsilon_{N,M}$ is defined in \eqref{eq:epsilon}.
\end{theorem}

We further examine three special cases in the following corollary, namely the multilayer SBM with independent layers \citep{lei2023bias}, the autoregressive SBM \citep{jiang2023autoregressive}, and the multilayer Ising model \citep{zhang2024}. These models correspond to the regimes $s_{kl}^*=0$, $s_{kl}^*\asymp M$, and $s_{kl}^*\asymp M^2$, respectively.

\begin{corollary}\label{Coro.compar}
    Suppose all assumptions in Theorem \ref{Thm.Comconsist} hold. When there exists no intra-layer dependence, $K \lesssim \log(NM)$ and $\eta_{N,M}\gtrsim \log(NM)$,
    \begin{enumerate}
         \item  if $s_{kl}^*\lesssim M$,  then $\operatorname{dist}(\widehat{\boZ},\boZ^*) = 
         O_p\left(\frac{1}{\rho_{N,M}}\left(\frac{1}{N^2} + \frac{1}{NM}\right)\right)
         $, up to some logarithmic terms;
        \item if $s_{kl}^*\asymp M^2$, then $\operatorname{dist}(\widehat{\boZ},\boZ^*) = 
         O_p\left(\frac{1}{\rho_{N,M}}\left(\frac{M}{N^2} + \frac{1}{NM}\right)\right)
         $, up to some logarithmic terms.
    \end{enumerate}
\end{corollary}

Corollary \ref{Coro.compar} guarantees that MLP consistently estimates the true community structures. This consistency holds for a diverging $K$, as long as it does not grow too rapidly. Under conditions of Corollary \ref{Coro.compar}, we make some comparisons between the proposed MLP and the existing methods. For the multilayer SBM with independent layers, Corollary \ref{Coro.compar} shows that MLP achieves an improvement over the result $O_{p}\left(\frac{1}{N^2} + \frac{1}{N^2 M \rho_{N,M}^2}\right)$  of \citet{lei2023bias} under the sparsity condition $\rho_{N,M} \ll \frac{1}{N+M}$. 
For the autoregressive SBM, the misclassification rate derived by \citet{jiang2023autoregressive} is $O_p\big( \frac{1}{M^2} + \frac{1}{N^2}\big)$ up to some logarithmic terms, under the sparsity condition $\rho_{N,M}\asymp 1$. In comparison, Corollary \ref{Coro.compar} implies that MLP yields a better rate than that in \citet{jiang2023autoregressive} when $\rho_{N,M} \gtrsim \frac{1}{\log(NM)}$ and $M \lesssim N$. Additionally,  Corollary \ref{Coro.compar} suggests that MLP can accommodate more general sparse scenarios, such as $\rho_{N,M} \asymp \frac{\log^2(NM)}{NM}$. For the multilayer Ising model, when $M\lesssim N$, the required sparsity condition in \citet{zhang2024} is that $\rho_{N,M} \gg \left( \frac{1}{N} \right)^{\frac{1}{1+c}}$ for some constant $c > 0$, up to some logarithmic terms. In contrast, Corollary \ref{Coro.compar} suggests that MLP can achieve 
$ \rho_{N,M} \gg \frac{1}{N} $, up to some logarithmic terms, which clearly achieves a better sparsity condition than that in \citet{zhang2024}.

\section{Numerical experiments}\label{Sec:experiments}
In this section, we evaluate the performance of MLP on both synthetic and real-world multilayer networks. We compare its performance against several widely used community detection methods in the literature, including the least squares estimation method (LSE; \citealp{lei2020consistent}), bias-adjusted spectral clustering (BASC; \citealp{lei2023bias}), multilayer Ising dependence (MID; \citealp{zhang2024}), and the co-regularized spectral clustering method (CRSC; \citealp{paul2020spectral}).

Let $\boe^* = \big(e_1^*, \dots, e_N^*\big)$ and $\widehat{\boe} = \big(\widehat{e}_1, \dots, \widehat{e}_N\big)$ represent the true and estimated community assignments, respectively. Note that the error metric in \eqref{eq:distZ} requires enumeration of all possible permutations, which is computationally infeasible for large networks. Therefore, we measure the community detection accuracy by an alternative metric \citep{wang2010consistent},
\begin{equation}\label{eq:Err}
    \operatorname{Err}(\boe^*, \widehat{\boe}) = \frac{2}{N(N-1)} \sum_{i<j} \mathbb{I}\left( \Big( \mathbb{I}(e_i^* = e_j^*) + \mathbb{I}(\widehat{e}_i = \widehat{e}_j) \Big) = 1 \right).
\end{equation}
which quantifies the number of mismatched pairs between $\boe^*$ and $\widehat{\boe}$.

\subsection{Synthetic networks}

To evaluate the performance of all competing methods, we consider seven simulated examples, each generated using distinct schemes for $\boTheta^*$ and $\boZ^*$. For illustration, we consider the sparse covariance scenario by setting $T_{kl}$ in the parameter space $\boOmega$ as the collection of the support of $\boSigma_{kl}^*$. Throughout Examples 1 - 6, the edge dependence structure $\boldsymbol{P}^*$ is specified as follows: for all $b\in[M]$, $r_{i_1 j_1,, i_2 j_2}^{(b)} = r_{i_1 j_1,, i_2 j_2} \sim \operatorname{Uniform}[-0.5, 0.5]$, with the correlation density for each edge fixed at 0.2.
 
\textbf{Example 1}. In this example, we consider the SBM with varying strengths of dependence. Let $N = 100$, $M = 20$, and $K = 5$. We first generate $\boZ^*$ using a multinomial distribution with parameters $(0.2, 0.2, 0.2, 0.2, 0.2)$.  Then, for each pair $(k, l)$, we define $\operatorname{diag}(\boSigma^{*}_{kl}) = \mathbf{1}_M$, with non-zero elements only in the first off-diagonal positions. Each of these off-diagonal elements is drawn from $\operatorname{Uniform}(-\sigma, \sigma)$ with $\sigma \in \{0, 0.2, 0.4, 0.6\}$ controlling the dependence strengths. In addition, we set the mean matrix elements $\mu_{kl}^{*,(b)} \sim N(-0.5, 0.01)\mathbb{I}\{k = l\} + N(-0.8, 0.01)\mathbb{I}\{k \neq l\}$.  

\textbf{Example 2}. In this example, we consider the SBM with varying network sparsity. Let $N = 100$, $M = 20$, and $K = 5$. The data generation scheme is similar to Example 1, and we set $\sigma = 0.2 $ to generate the first off-diagonal elements of $\boSigma^*_{kl}$. Then, for each $b < d$ and $k, l \in [K]$, we define  
$
\mu_{kl}^{*,(b)} \sim N(-v, 0.01)\mathbb{I}\{k = l\} + N(-v - 0.3, 0.01)\mathbb{I}\{k \neq l\},
$
where $v \in \{0.5, 0.8, 1.2\}$ controls the sparsity level of the network. 

\textbf{Example 3}. In this example, we consider the SBM with varying inter-layer dependence structures. Let $N = 100$, $M = 20$, and $K = 5$. We first generate the same $ \boZ^* $ as in Example 1 and set $v = 0.5$ as in Example 2 to generate $ \bomu^*$. To explore different inter-layer dependence structures, we vary $ s_{kl}^* $ by defining $ S_{a} = \{ (b,d) : |b - d| \leq a \} $, where $ a \in \{ 1, 2, 3, 4\}$ controls the extent of the dependence. Moreover, we set $\operatorname{diag}(\boSigma^{*}_{kl}) = \mathbf{1}_M$ and $\Sigma^{*,(bd)}_{kl} = \Sigma^{*,(bd)}_{lk}\sim N(\sigma-0.05|b-d|, 0.01)$ with $\sigma = 0.25$ for $b<d$ and $k,l\in[K]$. 

\textbf{Example 4}. In this example, we consider the SBM with varying network sizes. Let $M = 20$ and $K = 5$, and adopt the same generation scheme  for $\boTheta^*, \boZ^*$ as defined in Example 2 with $v = 0.5$. Additionally, we set $N = 80, 120, 160, 180$ to progressively increase the network size. 

\textbf{Example 5}. In this example, we consider the SBM with a varying number of layers. Let $N = 100$ and $K = 5$, and adopt the same generation scheme for $\boTheta^*,\boZ^*$ as defined in Example 2 with $v = 0.5$. Moreover, we set $M = 10,20,30,40$ to examine the effect of different layer counts. 

\textbf{Example 6}.  In this example, we consider the SBM with various community sizes. Let $N = 100, M = 20, K=5$. The data generating scheme for $\boTheta^*$ is the same as Example 2 with $v = 0.5$ except that we generate $\boZ^*$ differently. Specifically, we generate $\boZ^*$ using the multinomial distribution with the parameter being $(\gamma, \gamma, \gamma, 0.4-\gamma, 0.6-2\gamma)$. We vary $\gamma\in \{0.05,0.1,0.15\}$ and communities will become more unbalanced when $\gamma$ becomes smaller.

\textbf{Example 7}. 
In this example, we consider the SBM under different correlation densities.  Let $N = 100, M = 20, K=5$. The data-generating process for $\boTheta^*,\boZ^*$ follows the same setup as in Example 2, with $v = 0.5$. We let $  r_{i_1j_1, i_2j_2}^{(b)}= r_{i_1j_1, i_2j_2} \sim \operatorname{Uniform}[-0.2,0.2] $ for all $b\in[M]$, and vary the correlation density $\tau$ over the set $\{0, 0.3, 0.6\}$, where larger values of $\tau$ correspond to increasingly complex edge dependencies.

\textbf{Example 8}. In this example, we examine the robustness of MLP compared with other methods under a mis-specified model.  The data-generating process for $\boTheta^*,\boZ^*$ follows the same setup as in Example 2, with $v = 0.5$.   Let $N = 200, M = 8, K=5$.  Specifically, we consider a multilayer Ising model as described in Example 2 of \citet{zhang2024}, where $u\in \{-5, -5.5, -6\}$ controls the network sparsity, with the network becoming sparser as $u$ decreases.

\textbf{Example 9}.  In this example, we investigate the robustness of MLP under unknown support. Let $N = 100$, $M = 20$, $K = 5$, and set $r_{i_1j_1, i_2j_2}^{(b)} = r_{i_1j_1, i_2j_2} \sim \operatorname{Uniform}[-0.3, 0.3]$ for all $b \in [M]$, with $\tau = 0.2$. The data-generating mechanisms for $\bomu^*$ and $\boZ^*$ follow the same specification as in Example 2, with $v = 0.5$. To explore different inter-layer dependence structures, we vary $M\in \{20,25,30\}$ and set $s = s_{kl} =\lfloor M/2\rfloor$ by randomly selecting $s$ positions in the upper-triangular part of $\boSigma_{kl}$ and drawing their values from $\operatorname{Uniform}(-0.6, -0.2) \cup \operatorname{Uniform}(0.2, 0.6)$, while setting all remaining entries to zero. We apply the four-stage estimation procedure introduced in Section~\ref{Sec:Pairlike} to estimate community memberships by first estimating the support. For simplicity, we refer to the procedure involving support estimation as $\text{MLP}_{us}$, and denote the MLP with known support as $\text{MLP}_{ks}$.

Tables \ref{Table:sigma} - \ref{Table:s} summarize the averaged community detection errors defined in \eqref{eq:Err} along with their standard errors over 50 independent replications for all competing methods. It is evident that MLP consistently outperforms all competing approaches. This superior performance can be attributed to its ability to accurately capture the dependence structure as designed in all examples. In contrast, methods that do not account for this dependence fail to recover essential information, including the underlying community structure.

\begin{table}[!htb]
    \centering
    \caption{Averaged community detection errors (standard errors) over 50 independent replications for all competing methods in Example 1. The best performer in each case is bold-faced.}
    \label{Table:sigma}
    \begin{tabular}{cccccc}
        \toprule
        $\sigma$ & MLP  & LSE & CRSC & MID & BASC \\
        \midrule
        $0$   & \textbf{0.101} (0.058) & 0.289 (0.026) & 0.156 (0.034) & 0.295 (0.034) & 0.342 (0.018) \\
        $0.2$ & \textbf{0.107} (0.065) & 0.290 (0.021) & 0.162 (0.045) & 0.303 (0.034) & 0.340 (0.012) \\
        $0.4$ & \textbf{0.111} (0.067) & 0.292 (0.021) & 0.162 (0.042) & 0.298 (0.034) & 0.340 (0.015) \\
        $0.6$ & \textbf{0.098} (0.057) & 0.290 (0.023) & 0.160 (0.045) & 0.296 (0.032) & 0.341 (0.013) \\
        \bottomrule
    \end{tabular}
\end{table}

\begin{table}[H]
    \centering
    \caption{Averaged community detection errors (standard errors) over 50 independent replications for all competing methods in Example 2. The best performer in each case is bold-faced.}
    \label{Table:v}
    \begin{tabular}{cccccc}
        \toprule
        $v$ & MLP  & LSE & CRSC & MID & BASC \\
        \midrule
        $0.5$ & \textbf{0.047} (0.043) & 0.299 (0.016) & 0.115 (0.047) & 0.303 (0.021) & 0.326 (0.011) \\
        $0.8$ & \textbf{0.120} (0.061) & 0.303 (0.022) & 0.172 (0.042) & 0.302 (0.022) & 0.328 (0.011) \\
        $1.2$ & \textbf{0.229} (0.078) & 0.325 (0.029) & 0.237 (0.035) & 0.325 (0.029) & 0.330 (0.014) \\
        \bottomrule
    \end{tabular}
\end{table}

\begin{table}[H]
    \centering
    \caption{Averaged community detection errors (standard errors) over 50 independent replications for all competing methods in Example 3. The best performer in each case is bold-faced.}
    \label{Table:S}
    \begin{tabular}{cccccc}
        \toprule
        $S_a$ & MLP  & LSE & CRSC & MID & BASC \\
        \midrule
        $S_1$ & \textbf{0.104} (0.052) & 0.303 (0.015) & 0.154 (0.041) & 0.308 (0.022) & 0.327 (0.012) \\
        $S_2$ & \textbf{0.122} (0.065) & 0.288 (0.022) & 0.179 (0.045) & 0.299 (0.040) & 0.337 (0.017) \\
        $S_3$ & \textbf{0.128} (0.053) & 0.302 (0.015) & 0.171 (0.040) & 0.308 (0.020) & 0.326 (0.010) \\
        $S_4$ & \textbf{0.137} (0.052) & 0.302 (0.014) & 0.178 (0.036) & 0.308 (0.017) & 0.325 (0.011) \\
        \bottomrule
    \end{tabular}
\end{table}

\begin{table}[H]
    \centering
    \caption{Averaged community detection errors (standard errors) over 50 independent replications for all competing methods in Example 4. The best performer in each case is bold-faced.}
    \label{Table:N}
    \begin{tabular}{cccccc}
        \toprule
        $N$ & MLP  & LSE & CRSC & MID & BASC \\
        \midrule
        $80$ & \textbf{0.238} (0.045) & 0.312 (0.017) & 0.247 (0.030) & 0.312 (0.017) & 0.332 (0.016) \\
        $120$ & \textbf{0.040} (0.037) & 0.307 (0.014) & 0.093 (0.038) & 0.307 (0.015) & 0.324 (0.009) \\
        $160$ & \textbf{0.021} (0.037) & 0.048 (0.047) & 0.028 (0.039) & 0.049 (0.048) & 0.324 (0.008) \\
        $180$ & \textbf{0.001} (0.002) & 0.046 (0.052) & 0.006 (0.006) & 0.045 (0.050) & 0.323 (0.004) \\
        \bottomrule
    \end{tabular}
\end{table}

\begin{table}[H]
    \centering
    \caption{Averaged community detection errors (standard errors) over 50 independent replications for all competing methods in Example 5. The best performer in each case is bold-faced.}
    \label{Table:M}
    \begin{tabular}{cccccc}
        \toprule
        $M$ & MLP  & LSE & CRSC & MID & BASC \\
        \midrule
        $10$  & \textbf{0.210} (0.045) & 0.312 (0.009) & 0.232 (0.034) & 0.314 (0.013) & 0.328 (0.012) \\
        $20$ & \textbf{0.105} (0.064) & 0.290 (0.021) & 0.163 (0.045) & 0.303 (0.034) & 0.340 (0.012) \\
        $30$ & \textbf{0.014} (0.027) & 0.287 (0.024) & 0.046 (0.037) & 0.287 (0.024) & 0.328 (0.012) \\
        $40$ & \textbf{0.006} (0.018) & 0.268 (0.053) & 0.020 (0.024) & 0.271 (0.056) & 0.328 (0.011) \\
        \bottomrule
    \end{tabular}
\end{table}

\begin{table}[H]
    \centering
    \caption{Averaged community detection errors (standard errors) over 50 independent replications for all competing methods in Example 6. The best performer in each case is bold-faced.}
    \label{Table:gamma}
    \begin{tabular}{cccccc}
        \toprule
        $\gamma$ & MLP  & LSE & CRSC & MID & BASC \\
        \midrule
        $0.05$  & \textbf{0.205} (0.153) & 0.249 (0.061) & 0.320 (0.021) & 0.247 (0.061) & 0.444 (0.019) \\
        $0.10$  & \textbf{0.151} (0.106) & 0.231 (0.044) & 0.275 (0.023) & 0.229 (0.046) & 0.416 (0.025) \\
        $0.15$  & \textbf{0.091} (0.057) & 0.277 (0.023) & 0.164 (0.034) & 0.279 (0.026) & 0.341 (0.015) \\
        \bottomrule
    \end{tabular}
\end{table}

\begin{table}[H]
    \centering
    \caption{Averaged community detection errors (standard errors) over 50 independent replications for all competing methods in Example 7. The best performer in each case is bold-faced.}
    \label{Table:tau}
    \begin{tabular}{cccccc}
        \toprule
        $\tau$ & MLP  & LSE & CRSC & MID & BASC \\
        \midrule
        $0$  & \textbf{0.036} (0.050) & 0.295 (0.026) & 0.084 (0.048) & 0.306 (0.033) & 0.325 (0.008) \\
        $0.3$  & \textbf{0.040} (0.056) & 0.179 (0.074) & 0.041 (0.032) & 0.179 (0.074) & 0.329 (0.011) \\
        $0.6$  & \textbf{0.064} (0.052) & 0.308 (0.012) & 0.128 (0.050) & 0.313 (0.017) & 0.326 (0.009) \\
        \bottomrule
    \end{tabular}
\end{table}

\begin{table}[H]
    \centering
    \caption{Averaged community detection errors (standard errors) over 50 independent replications for all competing methods in Example 8. The best performer in each case is bold-faced.}
    \label{Table:u}
    \begin{tabular}{cccccc}
        \toprule
        $u$ & MLP  & LSE & CRSC & MID & BASC \\
        \midrule
        $-5$  & 0.102 (0.048) & 0.109 (0.016) & 0.196 (0.045) & \textbf{0.091} (0.010) & 0.237 (0.058) \\
        $-5.5$ & \textbf{0.209} (0.052) & 0.262 (0.002) & 0.302 (0.030) & 0.213 (0.009) & 0.259 (0.059) \\
        $-6$ & 0.399 (0.145) & 0.476 (0.054) & 0.412 (0.035) & \textbf{0.285} (0.015) & 0.490 (0.045) \\
        \bottomrule
    \end{tabular}
\end{table}
\begin{table}[H]
    \centering
    \caption{Averaged community detection errors (standard errors) over 50 independent replications for all competing methods in Example 9. The best performer in each case is bold-faced.}
    \label{Table:s}
    \begin{tabular}{ccccccc}
        \toprule
        $M$ & MLP$_{ks}$ & MLP$_{us}$ & LSE & CRSC & MID & BASC \\
        \midrule
        20 & \textbf{0.059} (0.045) & 0.061 (0.045) & 0.162 (0.058) & 0.085 (0.041) & 0.160 (0.059) & 0.346 (0.020) \\
        25 & \textbf{0.043} (0.042) & 0.046 (0.041) & 0.127 (0.057) & 0.054 (0.040) & 0.126 (0.057) & 0.351 (0.022) \\
        30 & \textbf{0.031} (0.035) & 0.037 (0.035) & 0.232 (0.047) & 0.083 (0.044) & 0.204 (0.067) & 0.335 (0.018) \\
        \bottomrule
    \end{tabular}
\end{table}

Tables \ref{Table:sigma} - \ref{Table:S} demonstrate that as $\sigma$, $v$, and $a$ increase, meaning that correlations among networks become more assertive, the networks become sparser, and the correlation structure becomes more complex, the performance of community detection for all methods declines. Nevertheless, MLP consistently achieves the best community detection accuracy. This highlights the advantage of fully considering the inter-layer dependence structure to improve the accuracy of community detection. Tables \ref{Table:N} and \ref{Table:M} show that as $N$ and $M$ increase, which corresponds to larger network sizes and a greater number of layers, the estimation errors of MLP, LSE, and CRSC all decrease. However, in comparison, MLP becomes increasingly stable, as both the mean error and standard deviation reach their minimum when $N = 180$ and $M = 40$. This suggests that as the amount of integrated information increases, the performance improvement of MLP becomes more evident.  Furthermore, Table~\ref{Table:gamma} indicates that as community imbalance increases, the difficulty of community detection also increases. Nevertheless, MLP continues to achieve the best performance among all approaches. Table \ref{Table:tau} further examines settings with more complex edge dependence, where the estimation errors of all methods become larger, but MLP continues to provide the most accurate estimates. Tables \ref{Table:u} and \ref{Table:s} further highlight the robustness of MLP against model mis-specification and unknown support. In addition, the comparison of computational cost for different methods in terms of both $M$ and $N$ is deferred to Appendix E.

\subsection{International trade network}

We apply the proposed MLP method to analyze an annual international trade network among $ N = 174 $ countries or regions from 1995 to 2014, with $ M = 20 $ layers. The dataset is a subset of the publicly available trade data covering 207 countries from 1870 to 2014 \citep{barbieri2009trading, barbieri2016cowtrade}. An edge between two countries or regions is defined as 1 if their total trade volume for the year falls within the top 20\% globally; otherwise, it is defined as 0. We then calculate the average edge density for each network layer, ranging from 0.126 in 1995 to 0.156 in 2014, which is moderately sparse. Furthermore, as suggested by \citet{lakdawala2025}, we set $ K = 4$ since all countries or regions can be grouped into four categories based on their global trade volumes.

\begin{figure}[!htb]
    \centering
    \begin{minipage}{0.49\linewidth}
        \centering
        \includegraphics[width=\linewidth]{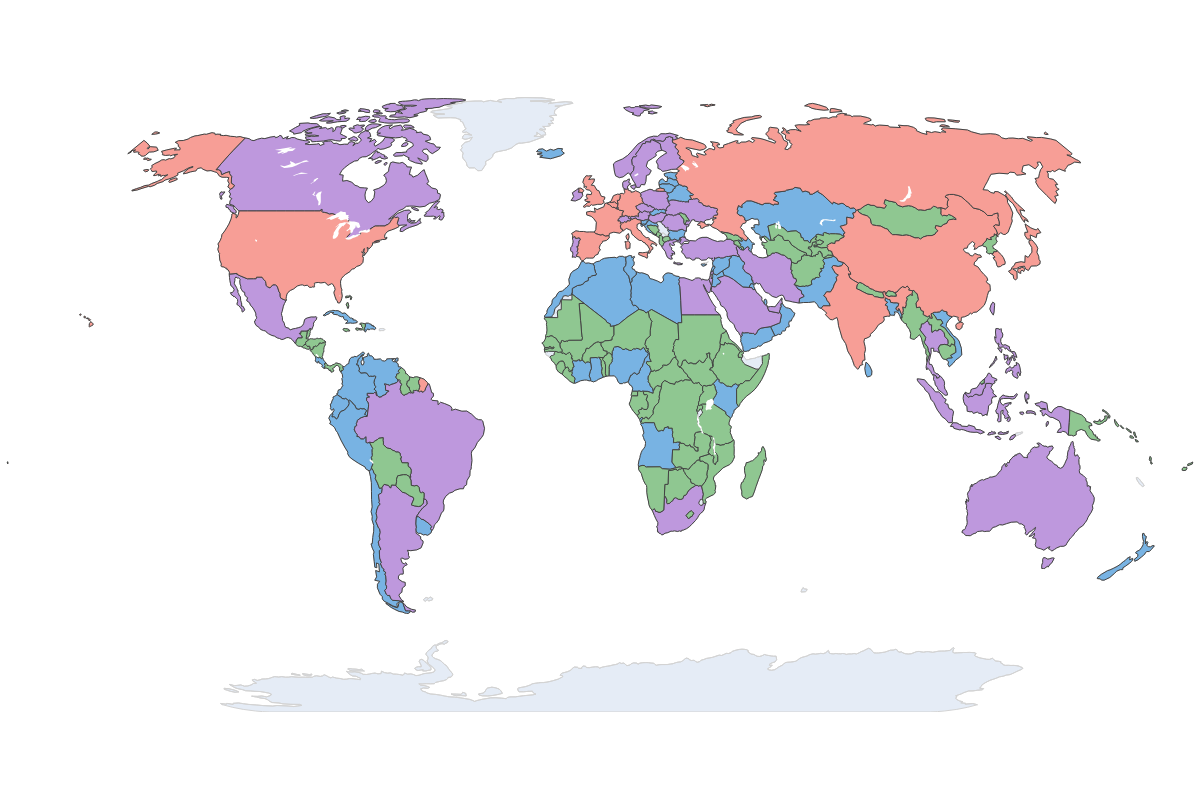}
        \caption*{LSE}
    \end{minipage}%
    \hspace{0.005\linewidth}
    \begin{minipage}{0.49\linewidth}
        \centering
        \includegraphics[width=\linewidth]{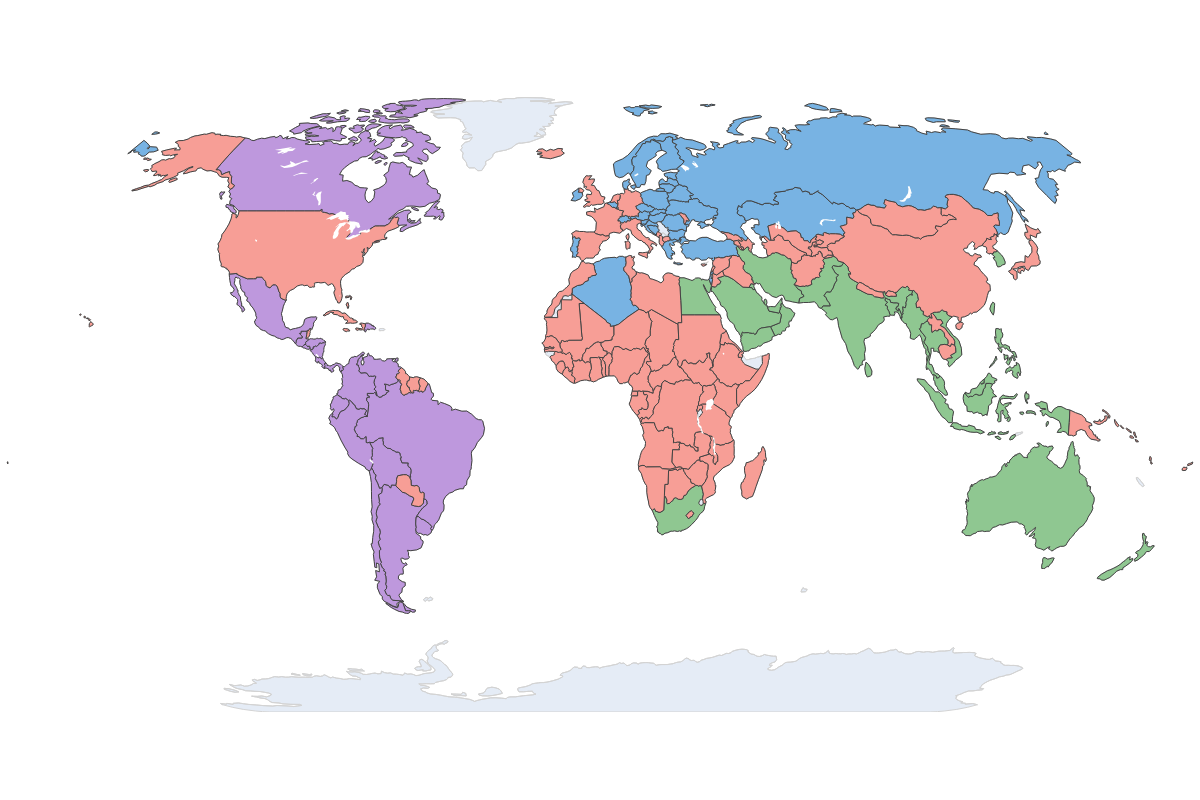}
        \caption*{CRSC }
    \end{minipage} \\

    \begin{minipage}{0.49\linewidth}
        \centering
        \includegraphics[width=\linewidth]{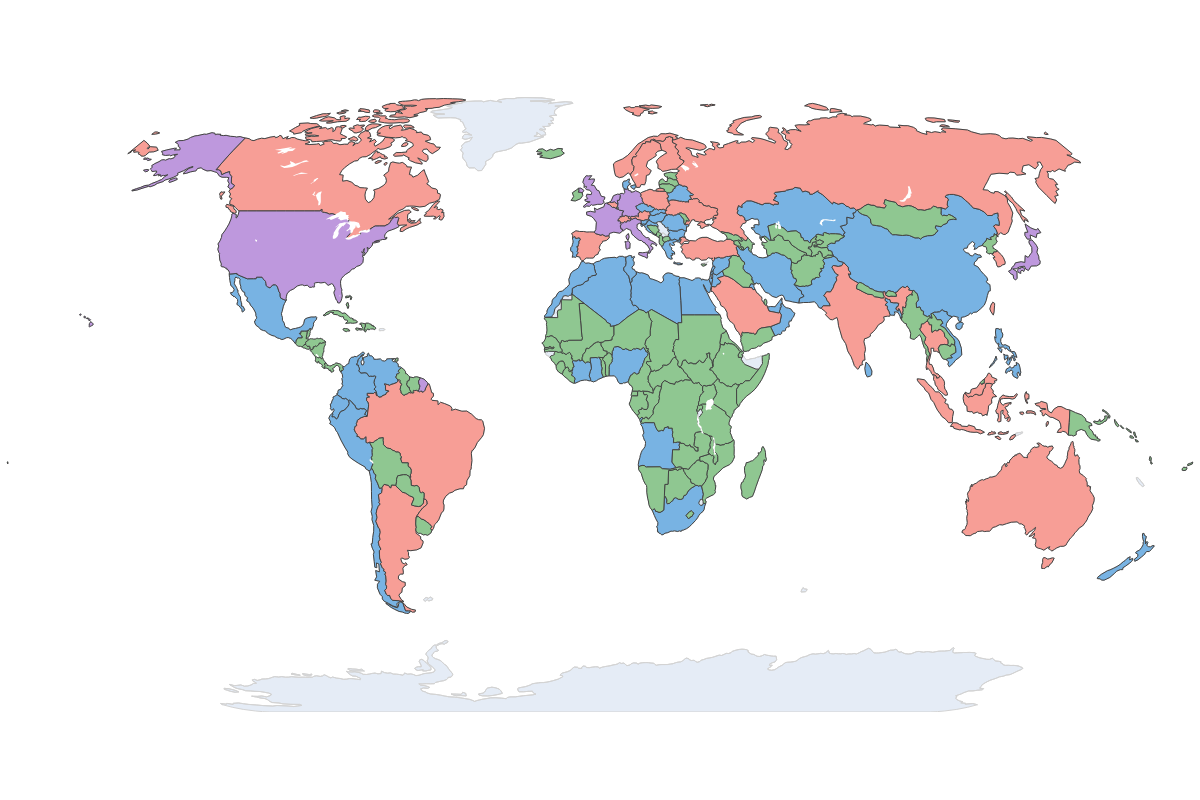}
        \caption*{MID }
    \end{minipage}%
    \hspace{0.005\linewidth}
    \begin{minipage}{0.49\linewidth}
        \centering
        \includegraphics[width=\linewidth]{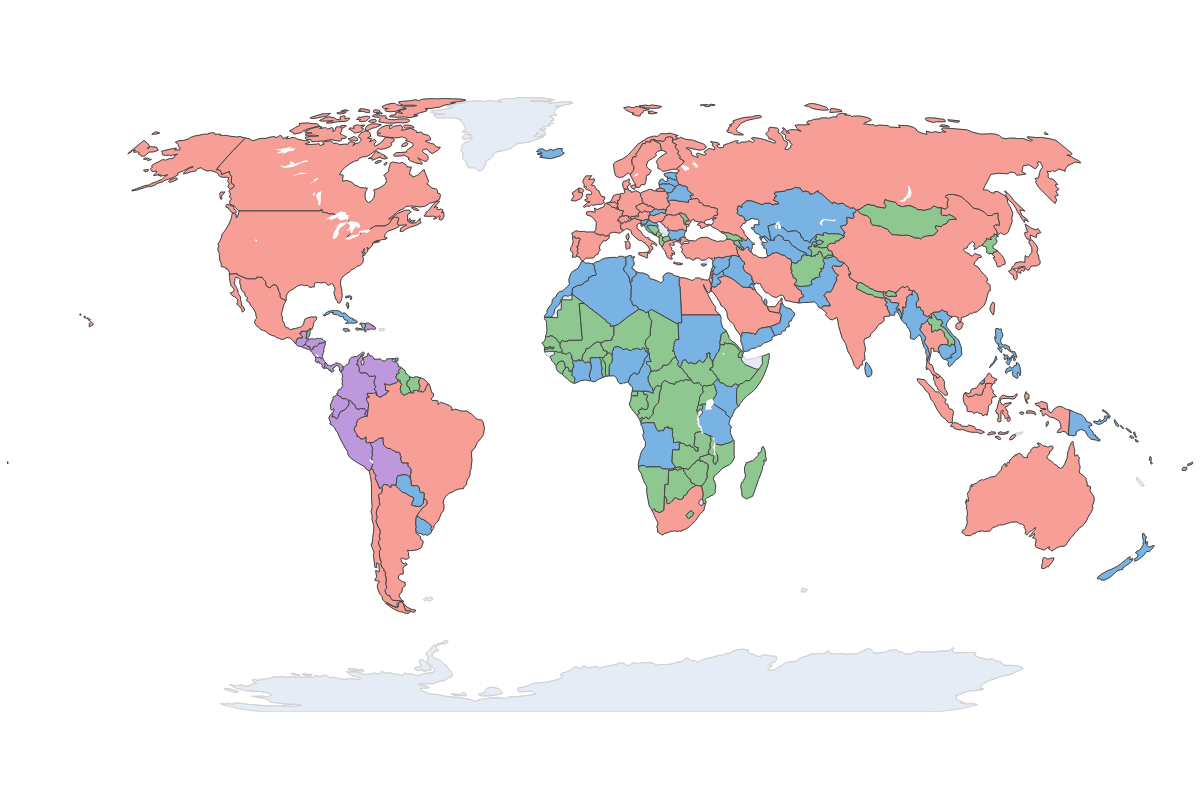}
        \caption*{MLP  }
    \end{minipage}

    \caption{The maps illustrate the results of community detection using different methods.}
    \label{fig:community}
\end{figure}

For illustration, we assume that only the adjacent network layers are dependent, and thus the dependence structure between different networks is the same as the one in Example 1 of Section 5.1. In fact, the trading networks corresponding to adjacent years exhibit strong correlations, as estimated by MLP, with Pearson correlation coefficients ranging from 0.8 to 0.95. 
We then apply the competing methods to detect community structure in the international trade network. The detailed community formations by different methods are provided in the Supplementary. Notably, BASC identifies four communities with sizes of 146, 30, 1, and 1, respectively. The highly unbalanced community sizes may not align with reality, and thus BASC is excluded from further discussion. Figure \ref{fig:community} visualizes the community structures detected by LSE, CRSC, MID, and MLP on the world map. 

It is evident that all four methods are consistent in identifying the community predominantly consisting of developing African countries and smaller economies, among which 66 countries or regions in common across all methods. These countries or regions generally have underdeveloped economies, and regional trade networks and tend to rely on specific trade partners. Furthermore, MLP is more accurate in identifying the world's major developed and emerging market economies. For example, between 2001 and 2014, key global economic cooperation organizations, such as APEC and G20, comprised core nations like the United States, China, Germany, India, Brazil, Canada, Australia, and Russia. The community colored red, as estimated by MLP, successfully recovers these crucial economies. In contrast, LSE, CRSC, and MID all place the United States, Canada, and Australia into different communities, failing to recognize their similar leading roles in the international trade network.

\section{Summary}\label{Sec:discuss}

In this article, we propose a multilayer probit network model which integrates the multilayer stochastic block model for community detection with
a multivariate probit model to capture the structures of inter-layer dependence, which
also allows intra-layer dependence.
%The key of the proposed method is to introduce a latent multivariate probit model to capture inter-layer and intra-layer dependence, whereas its covariance matrix assumes an SBM model for community detection. 
A constrained pairwise likelihood approach, paired with an efficient alternating updating algorithm, is developed to estimate the model parameters. Furthermore, asymptotic analysis, focusing on both parameter estimation and community detection, is conducted to ensure the desirable theoretical properties of the proposed method. 
These properties are further validated through numerical experiments on several simulated examples and a real-world multilayer trade network.

\section*{Acknowledgment}
H. Zhang acknowledges National Natural Science Foundation of China Grant 12501364.
T. Wang gratefully acknowledges 
Science and Technology Commission of Shanghai Municipality Grant 23JC1400700 and National Natural Science Foundation of China Grant 12301660. JW's research is supported in part by GRF-11311022, GRF-14306523, GRF-14303424, and CUHK Startup Grant 4937091.

\bibliographystyle{apalike}
\bibliography{ref}

\end{document}